\documentclass[aps,prl, reprint,superscriptaddress, preprintnumbers]{revtex4-1}
\usepackage{blindtext}
\usepackage[utf8]{inputenc}
\usepackage[english]{babel}
\usepackage{amsmath}

 \def\be{\begin{equation}}
\def\ee{\end{equation}}
 \def\ba{\begin{align}}
\def\ea{\end{align}}
\def\bea{\begin{eqnarray}}
\def\eea{\end{eqnarray}}
\def\pd{\partial}
\def\a{\alpha}
\def\b{\beta}

\def\m{\mu}
\def\n{\nu}

\def\l{\lambda}

\def\r{\rho}
\def\s{\sigma}

\def\D{\nabla}

\def\f{\phi}

\begin{document}

\preprint{IFT-UAM/CSIC-15-039}
\preprint{FTUAM-15-11}
\preprint{FTI/UCM 43/2015}

\title{Unimodular Gravity Redux}
\author{E. Álvarez}
\email{enrique.alvarez@uam.es}
\affiliation{Instituto de Física Teórica UAM/CSIC, C/ Nicolas Cabrera, 13–15, C.U. Cantoblanco, 28049 Madrid, Spain}
\affiliation{Departamento de Física Teórica, Universidad Autónoma de Madrid, 20849 Madrid, Spain}
\author{S. González-Martín}
\email{sergio.gonzalez.martin@csic.es}
\affiliation{Instituto de Física Teórica UAM/CSIC, C/ Nicolas Cabrera, 13–15, C.U. Cantoblanco, 28049 Madrid, Spain}
\affiliation{Departamento de Física Teórica, Universidad Autónoma de Madrid, 20849 Madrid, Spain}
\author{M. Herrero-Valea}
\email{mario.herrero@csic.es}
\affiliation{Instituto de Física Teórica UAM/CSIC, C/ Nicolas Cabrera, 13–15, C.U. Cantoblanco, 28049 Madrid, Spain}
\affiliation{Departamento de Física Teórica, Universidad Autónoma de Madrid, 20849 Madrid, Spain}
\author{C. P. Martín}
\email{carmelop@fis.ucm.es}
\affiliation{Universidad Complutense de Madrid (UCM), Departamento de Física Teórica I, Facultad de Ciencias Físicas,  Av. Complutense S/N (Ciudad Univ.), 28040 Madrid, Spain}
\begin{abstract}
 It is well known that the problem of the cosmological constant appears in a new light in Unimodular Gravity. In particular, the zero momentum piece of the potential does not automatically produce a corresponding cosmological constant. Here we show that quantum corrections do not renormalize the classical value of this observable.
\end{abstract}

\maketitle
\section{Introduction}
There are many facets to the problem of the cosmological constant. One of them is to explain why the vacuum energy do not produce a huge value for it. It would appear that either Wilsonian ideas of effective theories do not work in this case, or else that vacuum energy does not obey the equivalence principle.
\par
In Unimodular Gravity the vacuum energy (actually all potential energy) is naively decoupled from gravitation, because when the spacetime metric is unimodular, that is
\be
\hat{g}\equiv \text{det}~\hat{g}_{\m\n}=-1
\ee
the interaction term between the potential energy and the metric is of the type
\be
S_\text{in}\equiv \int d^n x \left({1\over 2}\hat{g}^{\m\n}\pd_\m\psi\pd_\n\psi- V(\psi)\right)
\ee
(where $\psi$ generically stands for any matter field), so that the matter potential does not couple to gravitation at the lagrangian level.
Actually things are not so simple though, and there is an interaction forced upon us by the Bianchi identity.
A novel aspect is that the unimodular condition breaks full diffeomorphism invariance $\text{Diff}(M)$ (where $M$ is the space-time manifold) to a subgroup $\text{TDiff}(M)$ consisting on those diffeomorphisms $x\rightarrow y$  that have unit jacobian, that is
\be
\text{det}{\pd y^\a\over \pd x^\m}=1
\ee
Incidentally, van der Bij,~van Dam and ~Ng \cite{vanderBij} showed a long time ago that $\text{TDiff}$  is enough to make gauge artifacts of the three excess gauge polarizations when going to the massless limit in a spin two flat space theory. (There are five polarizations in the massive case and only two in the massless limit).
\par

It is possible (and technically convenient) to formulate the theory in such a way that it has an added Weyl invariance by writing
\be
\hat{g}_{\m\n}\equiv g^{-{1\over n}}~g_{\m\n}
\ee
The reason is that then the variations $\delta g_{\a\b}$ are unconstrained, whereas
\be
\hat{g}^{\a\b}\delta \hat{g}_{\a\b}=0
\ee
We shall first recall in what sense the issue of the cosmological constant is changed in Unimodular Gravity; this  we first do in a flat space setting, and then in the full nonlinear theory. We then will report on a one-loop calculation in which we computed the counterterm and found that no cosmological constant is generated to this order. 
The argument can actually be extended  to any order in perturbation theory.

\section{Flat space gauge symmetry}
In reference \cite{ABGV} an analysis of the most general action principle built out of dimension four operators for a spin two field $h_{\m\n}$  was made, namely
\be
L\equiv \sum_{i=1}^4 C_i {\cal O}^{(i)}
\ee
\bea
&&{\cal O}^{(1)}\equiv {1\over 4}\pd_\m h_{\r\s}\pd^\m h^{\r\s}\nonumber\\
&&{\cal O}^{(2)}\equiv -{1\over 2}\pd^\r h_{\r\s}\pd_\m h^{\m\s}\nonumber\\
&&{\cal O}^{(3)}\equiv {1\over 2}\pd_\m h\pd_\l h^{\m\l}\nonumber\\
&&{\cal O}^{(4)}\equiv -{1\over 4}\pd_\m h\pd^\m h
\eea
where all indices are raised and lowered with the flat space metric $\eta_{\m\n}$, and $h\equiv \eta^{\m\n} h_{\m\n}$.
Also $C_1=1$ fixes the global normalization.
The result of the analysis was that $\text{LTDiff}$ (that is, linearized $\text{TDiff}$) invariance forces
\be
C_2=1
\ee
where $\text{LTDiff}$ symmetry is just invariance under the transformations
\be
\delta h_{\m\n}=\pd_\m \xi_\n+\pd_\n\xi_\m
\ee
with
\be
\pd_\m\xi^\m=0
\ee
The most important result was however the following. Amongst all the $\text{TDiff}$ invariant theories obtained for arbitrary values of $C_3$ and $C_4$ there are only two that propagate spin two only, without any admixture of spin zero. Those are
\be
C_3=C_4=1
\ee
which has an enhanced symmetry under linearized diffeomorphisms (without the transversality restriction). This is the Fierz-Pauli theory.
\par
The other one corresponds to
\bea
&&C_3={2\over n}\nonumber\\
&&C_4={n+2\over n^2}
\eea
This second theory is actually a truncation  of the Fierz-Pauli one, obtained by
\be
h_{\m\n}\rightarrow h_{\m\n}-{1\over n} h \eta_{\m\n}
\ee
(which is not a field redefinition, because it is not invertible). This theory was called $\text{WTDiff}$ and is actually the linear limit of Unimodular Gravity.

\section{The non-linear regime}
The truncation of General Relativity to unimodular metrics is simply
\small
\bea
&&S\equiv - M_P^{n-2}\int d^n x \left(R[\hat{g}]+L_{\text{matt}} [\psi_i,\hat{g}]\right)=\\
\nonumber &&-M_P^{n-2}\int d^n x ~ |g|^{1\over n}~\left(R+{(n-1)(n-2)\over 4 n^2}{\nabla_\m g\nabla^\m g\over g^2}+ L_{\text{matt}}\right)
\eea
\normalsize

In terms of a general metric, the equations of motion (EM) are given \cite{ABGV} by the manifestly traceless expression
\begin{widetext}
\begin{align}\label{traceless_EE}
R_{\m\n}-\frac{1}{n}R g_{\m\n}+{(n-2)(2n-1)\over 4 n^2} \left({\nabla_\m g\nabla_\n g \over g^2}-{1\over n} {(\nabla g)^2\over g^2} g_{\m\n}\right)-{n-2\over 2n} \left({\nabla_\m \nabla_\n g \over g}-{1\over n} 
{\nabla^2 g\over g^2} g_{\m\n}\right)=M_P^{2-n}\left(T_{\m\n}-\frac{1}{n}T g_{\m\n}\right)
\end{align}
\end{widetext}
When $|g|=1$ they are quite similar to the ones posited in 1919 by Einstein for obscure reasons \cite{Einstein}\cite{Ellis} related to Mie's theory.

Now it is easy to see that the Bianchi identities bring the trace back into the game, albeit in a slightly different form
\begin{align}
\frac{n-2}{2n}\D_{\m}R=-\frac{1}{n}\D_{\m}T
\end{align}
which integrates to
\begin{align}
\frac{n-2}{2n}R+\frac{1}{n}T=-C
\end{align}
so that going back to \eqref{traceless_EE} we recover the full Einstein equations
\begin{align}
R_{\m\n}-\frac{1}{2}R g_{\m\n}-C g_{\m\n}=T_{\m\n}
\end{align}
with an arbitrary integration constant which takes the role of a cosmological constant and whose value is to be defined by boundary conditions. Moreover, since the original equations couple to the traceless part of the energy-momentum tensor, any possible vacuum energy, or dynamical cosmological constant coming from a non-trivial minimum in the potential of a scalar field, is absorbed into the still arbitrary constant.
\section{Quantum corrections}
We have computed \cite{AGMHVM} the quantum corrections around an arbitrary unimodular background
\be
g_{\m\n}\equiv \bar{g}_{\m\n}+\kappa h_{\m\n}
\ee
with
\be
|\bar{g}|=1
\ee
The calculation has some technical complications stemming from the fact that the  full gauge symmetry is a combination of $\text{TDiff}(M)$ with $\text{Weyl}(M)$. This makes the BRST gauge fixing somewhat involved.

This makes the gauge fixing somewhat involved and solving the problem requires the introduction of a gauge fixing term through a BRST quantisation instead of the usual and simpler Faddeev-Poppov technique. The most important issue is that we need to introduce a new collection of fields representing not only the usual Faddeev-Poppov ghost fields but also a collection of Nielsen-Kallosh ghosts in order to implement the constraint over transverse diffeomorphisms $\D_{\m}c^{T\m}=0$. The full set of fields required for the quantization is then
\begin{align*}
&h^{(0,0)}_{\m\n},\; c^{(1,1)}_{\m},\; b_{\m}^{(1,-1)},\; f_{\m}^{(0,0)},\; \f^{(0,2)}\\
&\pi^{(1,-1)},\; \pi^{'(1,1)},\; \bar{c}^{(0,-2)},\; c^{'(0,0)}\\
&c^{(1,1)},\; b^{(1,-1)},\;f^{(0,0)},
\end{align*}
where $(m,n)$ are the Grassman number, defined mod 2, and ghost number of each field.

This way $S_{UG}+S_{fixing}$ will then contain interactions between all the different fields. However, by choosing the gauge fixing properly, all the complications can be contained in only one of the quadratic operators, the one involving $h_{\m\n}$, $f$ and $c'$ which happens to be non-minimal. Its determinant has been computed using the generalized Barvinsky-Vilkovisky technique developed by \cite{Barvinsky}. Full details will be given in \cite{AGMHVM} but let us emphasize that we have been able to keep one free gauge parameter in all intermediate steps, which should cancel when putting the final counterterm on shell owing to Kallosh theorem. It does indeed, which is a test of our computations. 

\par
At any rate the free background EM are given by
\be
\bar{R}_{\m\n}={1\over n}~\bar{R}\bar{g}_{\m\n}
\ee
which implies
\be
\bar{R}=\text{constant}
\ee
as well as
\be
\bar{R}_{\m\n}\bar{R}^{\m\n}={1\over n} \bar{R}^2=\text{constant}
\ee
This in turn means that the Weyl tensor squared $W_4$, which is related to the Euler density $E_4$ by the four-dimensional relationship
\be
W_4=E_4+2 \bar{R}_{\m\n} \bar{R}^{\m\n}-{2\over 3}\bar{R}^2=E_4+\text{constant}
\ee
is also a topological density modulo a dynamically irrelevant term.
\par
The one loop counterterm is 
\begin{align}
\nonumber &S_\infty= {1\over 16\pi^2 (n-4)}\int d^4x\left(\dfrac{119}{90}R_{\m\n\r\s}R^{\m\n\r\s} -\dfrac{83}{120}R^2\right)=\\
&={1\over 16\pi^2 (n-4)}\int d^4 x \left(\frac{119}{90  }E_4+\text{constant}\right)
\end{align}
This is to be contrasted with the counterterm of General Relativity with a cosmological constant \cite{Christensen:1979iy}

\be
S_{GR}={1\over 16 \pi^2 \left(n-4\right)}\int \sqrt{|\bar{g}|}~d^4 x \left(-{1142\over 135}\Lambda^2+{53\over 45} W_4\right)
\ee

Now the important thing is that all constant contributions which usually contribute to the cosmological constant because they couple to the metric through the volume element, are here dynamically irrelevant precisely because they do not couple to the gravitational field. Those terms that we computed, in spite of being gauge independent are then physically irrelevant.

\section{Conclusions}
It has been argued in this paper that quantum corrections do not generate a cosmological constant in Unimodular Gravity. It would be more precise to say that the cosmological constant is generated, but it does not couple to the gravitational field.

It is worth remarking that, although we have performed an explicit computation at the one-loop level only, our result can be extended to any loop order, since it relies only in the fact that the corresponding operators are non-dynamical (or in the absence of anomaly) and not in the particular numerical value of the counterterms.

The result reported in this paper is not a consequence of having chosen $\bar{g}=-1$, but rather it stems from Weyl symmetry which prevents dimension zero terms in the action. That is
\be
S=\int d^n x \left(-\bar{g}\right)^\b
\ee
for any nonvanishing $\b$.
\par

We believe this is a step forward in the understanding of the cosmological constant.

Unimodular Gravity is quite close to General Relativity in spite of the technical complications caused by the absence of full $\text{Diff}(M)$ symmetry. It is worth exploring it in further detail to fully understand the subtle differences between both theories.

\section{Acknowledgments}
We acknowledge useful discussions with A. O. Barvinsky and C. F. Steinwachs. This work has been partially supported by the European Union FP7  ITN INVISIBLES (Marie Curie Actions, PITN- GA-2011- 289442)and (HPRN-CT-200-00148) as well as by FPA2012-31880 (MICINN, Spain)), FPA2011-24568 (MICINN, Spain), S2009ESP-1473 (CA Madrid) and COST Action MP1210 (The String Theory Universe). The authors acknowledge the support of the Spanish MINECO {\em Centro de Excelencia Severo Ochoa} Programme under grant  SEV-2012-0249. 



\end{document}